\documentclass{article}

     \PassOptionsToPackage{numbers, compress}{natbib}

\usepackage[preprint]{neurips_2023}

\usepackage[utf8]{inputenc} 
\usepackage[T1]{fontenc}    
\usepackage{hyperref}       
\usepackage{url}            
\usepackage{booktabs}       
\usepackage{amsfonts}       
\usepackage{nicefrac}       
\usepackage{microtype}      
\usepackage{xcolor}         
\usepackage{amsmath}
\usepackage{amssymb}
\usepackage{mathtools}
\usepackage{amsthm}
\usepackage{multirow}
\usepackage{wrapfig}
\usepackage[capitalize,noabbrev]{cleveref}
\usepackage{graphicx}
\usepackage{subfigure} 
\usepackage{float}
\usepackage{algorithm} 
\usepackage{algorithmic}
\usepackage{textcomp} 
\usepackage{listings} 
\usepackage{enumitem}

\title{Make-An-Audio 2: Temporal-Enhanced \\ Text-to-Audio Generation}

\author{
    Jiawei Huang\thanks{Equal contribution.} \thanks{Interns at ByteDance.}  \\
    Zhejiang University, ByteDance\\
    \texttt{huangjw@zju.edu.cn} \\
    \And
    Yi Ren\footnotemark[1]  \\
    ByteDance\\  
    \texttt{ren.yi@bytedance.com} \\
    \And
    Rongjie Huang \\
    Zhejiang University\\
    \texttt{rongjiehuang@zju.edu.cn} \\
    \And
    Dongchao Yang \\
    Peking University\\
    \texttt{2001212832@stu.pku.edu.cn} \\
    \And
    Zhenhui Ye\footnotemark[2] \\
    Zhejiang University, ByteDance\\
    \texttt{zhenhuiye@zju.edu.cn} \\
    \And
    Chen Zhang, Jinglin Liu, Xiang Yin, Zejun Ma \\
    ByteDance\\
    \texttt{\{zhangchen.0620,jinglinliu,yinxiang.stephen,mazejun\}@bytedance.com} \\
    \And
    Zhou Zhao\thanks{Corresponding author} \\
    Zhejiang University \\
    \texttt{zhaozhou@zju.edu.cn} \\
}

\begin{document}

\maketitle
\begin{abstract}
Large diffusion models have been successful in text-to-audio (T2A) synthesis tasks, but they often suffer from common issues such as semantic misalignment and poor temporal consistency due to limited natural language understanding and data scarcity. Additionally, 2D spatial structures widely used in T2A works lead to unsatisfactory audio quality when generating variable-length audio samples since they do not adequately prioritize temporal information. To address these challenges, we propose Make-an-Audio 2, a latent diffusion-based T2A method that builds on the success of Make-an-Audio. Our approach includes several techniques to improve semantic alignment and temporal consistency: Firstly, we use pre-trained large language models (LLMs) to parse the text into structured <event \& order> pairs for better temporal information capture. We also introduce another structured-text encoder to aid in learning semantic alignment during the diffusion denoising process. To improve the performance of variable length generation and enhance the temporal information extraction, we design a feed-forward Transformer-based diffusion denoiser. Finally, we use LLMs to augment and transform a large amount of audio-label data into audio-text datasets to alleviate the problem of scarcity of temporal data. Extensive experiments show that our method outperforms baseline models in both objective and subjective metrics, and achieves significant gains in temporal information understanding, semantic consistency, and sound quality.
\end{abstract}

\section{Introduction}
Deep generative learning models~\cite{goodfellow2020generative,kingma2018glow,ho2020denoising} have revolutionized the creation of digital content, enabling creators with no professional training to produce high-quality images 
~\cite{rombach2022high,saharia2022photorealistic,nicholGLIDEPhotorealisticImage2021}, vivid videos~\cite{hong2022cogvideo,singer2022make}, diverse styles of voice~\cite{huang2022generspeech}, and meaningful long textual spans~\cite{zhang2022opt,openai2023gpt4}. 
Professional practitioners can modify the generated content to accelerate their production workflows. Text-to-audio synthesis (T2A) is a subcategory of generative tasks that aims to generate natural and accurate audio by taking text prompts as input. T2A can be useful in generating desired sound effects, music or speech, and can be applied to various applications like movie sound effects making, virtual reality, game development, and audio editing.

Thanks to the development of text-to-image synthesis (T2I) methods, researchers have successfully extended similar approaches to the text-to-audio synthesis domain~\cite{huang2023makeanaudio,liuAudioLDMTexttoAudioGeneration2023,yangDiffsoundDiscreteDiffusion2023,kreukAudioGenTextuallyGuided2023}. The success of these methods has opened up numerous opportunities for generating high-quality audio content from text.
T2A systems typically use a text encoder to encode the audio's text input as condition embedding, then employ diffusion models~\cite{huang2023makeanaudio,liuAudioLDMTexttoAudioGeneration2023,yangDiffsoundDiscreteDiffusion2023} to synthesis mel-spectrograms or utilize autoregressive models~\cite{kreukAudioGenTextuallyGuided2023} to synthesis raw waveform data based on the condition embedding. However, previous T2A methods have some common issues: 1) \textbf{Temporal disorder}: when the text input is complex, with multiple objects and temporal relationships between them, the generated audios often suffer from semantic misalignment and temporal disorder. For instance, audio captions such as "The sound of A, followed by the sound of B" may result in audios where A and B overlapping throughout, or B comes before A, or even only one sound is synthesized. 2) \textbf{Poor variable-length results}: previous works~\cite{huang2023makeanaudio} adopt the conventional U-Net structure of 2D convolution and spatial transformer stacking as the backbone of diffusion denoiser, which is typically trained with fixed-length audios. Consequently, they generate suboptimal results when synthesizing audio sequences of varying lengths compared to those of the training data. In the meanwhile, 2D spatial structures are not good at extracting temporal information since they treat the time axis and frequency axis equally in the spectrogram generation process. 3) \textbf{Insufficient temporal paired data}: previous works use simple rule-based augmentation methods~\cite{elizalde2022clap,kreukAudioGenTextuallyGuided2023} to create temporally aligned text-audio paired data from audio-label datasets. However, these patterns are overly simplistic and can hinder the model's ability to generalize to real-world sentences.

In this paper, based on previous successful work Make-An-Audio, we propose a novel temporal-enhanced text-to-audio generation framework called Make-An-Audio 2. The temporal information can be better handled by our method in the following ways: 1) To address the semantic misalignment and temporal disorder, we use a pre-trained LLM to extract the audio caption's temporal information and parse the origin caption into structured <event \& order> pairs with proper prompts. To encode the structured pairs better, we introduce another structured-text encoder that takes the structured pairs as its input to aid in learning semantic alignment during the diffusion denoising process. In this way, we relieve the text encoder's burden of recognizing events with the corresponding temporal information and enable the T2A system to model the timing information of the events more effectively. 2) To improve the generation quality of variable-length audio and enhance the temporal information understanding, we replace the 2D spatial structures with temporal feed-forward Transformer~\cite{vaswani2017attention} and 1D-convolution stacks for the diffusion denoiser and support variable-length audio input in training. 3) To address the issue of insufficient temporally aligned audio-text paired dataset, we use single-labeled audio samples and their labels to compose complex audio and structured captions. We then use LLM to augment the structured caption into natural language captions. 

We conduct extensive experiments on AudioCaps and Clotho datasets, which reveals that our method surpasses baseline models in both objective and subjective metrics, and achieves significant gains in understanding temporal information, maintaining semantic consistency, and enhancing sound quality. Our ablation studies further demonstrate the effectiveness of each of our techniques\footnote{Our demos are available at \url{https://make-an-audio-2.github.io/}}.

\section{Related works}
\subsection{Text-to-image generative models}
Text-to-Image Synthesis (T2I) has garnered significant attention in recent years and has even been commercialized. One pioneering work in this realm is DALL-E~\cite{ramesh2021zero}, which treats T2I generation as a sequence-to-sequence translation task. DALL-E employs a pre-trained VQ-VAE~\cite{van2017neural} to encode image patches to discrete codes, which are then combined with the text codes. During inference, the model generates image codes autoregressively based on the text codes. As diffusion models exhibit greater potential with regard to both diversity and quality in image generation, they have become mainstream in T2I Synthesis. DALLE-2~\cite{ramesh2022hierarchical} uses the CLIP~\cite{radford2021learning} text encoder and two diffusion models. The first diffusion model predicts CLIP visual features based on the CLIP text feature, while the second synthesizes the image from the predicted CLIP visual features. A cascade of diffusion super-resolution models is then employed to increase the resolution of the generated image. Another famous T2I work is Imagen~\cite{saharia2022photorealistic}, which utilizes the T5 encoder~\cite{raffel2020exploring} to extract text features, It employs a diffusion model to synthesize a low-resolution image and then applies a cascade of diffusion models for super-resolution. Latent Diffusion~\cite{rombach2022high} enhances computational efficiency by using a continuous VAE that is trained with a discriminator to map images from pixel space to compressed latent space. This is followed by diffusion on the latent space, which synthesizes images' latent.

\subsection{Text-to-audio synthesis}
Text-to-Audio Synthesis is a rising task that has seen great advances recently. Diffsound~\cite{yangDiffsoundDiscreteDiffusion2023} uses a pre-trained VQ-VAE~\cite{van2017neural} trained on mel-spectrograms to convert audio into discrete codes, which are then used by a diffusion model to generate the audio codes. To improve its generalization ability, the authors pre-trained Diffsound on the AudioSet dataset, which contains audio files labeled with tags. Additionally, they introduce a random input masking technique to make use of these tags. AudioGen~\cite{kreukAudioGenTextuallyGuided2023} is another system in this field that uses a similar VQ-VAE-based approach. However, it encodes raw waveform data into discrete codes and employs an autoregressive model to predict audio tokens based on text features. For data augmentation, AudioGen mixes audio files and concatenates their text captions.
Make-An-Audio~\cite{huang2023makeanaudio}, AudioLDM~\cite{liuAudioLDMTexttoAudioGeneration2023}, and TANGO~\cite{ghosal2023texttoaudio} are all based on the Latent Diffusion Model (LDM). With the assumption that CLAP can map the audio and its caption to the same latent space and approximate the text features based on the audio feature, AudioLDM uses audio features extracted by the CLAP model as the condition during training and utilizes text features during inference. In contrast, Make-An-Audio and TANGO employ text features both in the training and inference stages. To overcome data scarcity, Make-An-Audio proposes a pseudo prompt enhancement method, while TANGO introduces an audio mixture method based on human auditory perception.

\subsection{LLM-based data augmentation}
Recent advancements in prompt learning~\cite{liu2021pretrain} have greatly enhanced the capabilities of language models and given birth to very large language models with billions of parameters, enabling them to achieve natural language comprehension levels that are comparable to those of humans. OPT~\cite{zhang2022opt}, ChatGPT\footnote[1]{\url{https://openai.com/blog/chatgpt}} and GPT4~\cite{openai2023gpt4}, are typical cases among them. This has led researchers to explore whether these models can be used to annotate data instead of humans. In previous work, masked language models (MLM) like BERT~\cite{devlin2018bert} and Roberta~\cite{liu2019roberta} have been used for contextual augmentation~\cite{kumar2021data} at the word level. For example, researchers insert <mask> tokens into the text or replace some words with <mask> tokens, and then use the MLM to predict the appropriate words. At the sentence level, back translation~\cite{sennrich2015improving} and paraphrasing~\cite{kumar2019submodular} methods have been used to increase the diversity of data. However, the limited capabilities of these models have resulted in insufficient quality and diversity of the generated data.
To address these limitations, recent research has explored the use of very large language models for data augmentation. AugGPT~\cite{dai2023auggpt}, for instance, leverages ChatGPT to generate auxiliary samples for few-shot text classification. The quality of the generated data is much higher, resulting in double-digit improvements in sentence classification accuracy compared to previous data augmentation methods. WavCaps~\cite{mei2023wavcaps} crawls audio data with raw descriptions from multiple web sources. However, the raw descriptions contain a high degree of noise, so ChatGPT is used to filter out extraneous information unrelated to the audio and generate high-quality captions based on labels. This approach results in a large-scale, weakly-labeled audio captioning dataset.

\section{Preliminary}
Our work is inspired by Make-An-Audio~\cite{huang2023makeanaudio}, which is one of the most successful T2A models. In this section, we provide a preliminary discussion on Make-An-Audio. Denote an audio-text pair as $(a,y)$ where $a \in R^{T_a}$ and $T_a$ is the waveform length. To mitigate the complexity of modeling long continuous waveform data, they first convert $a$ to mel-spectrogram (akin to 1-channel 2D images) $x \in R^{C_a \times T}$, where $C_a, T \ll T^a$ denote the mel-channels and the number of frames respectively. The training process includes two stages:
  
1) \textbf{Training variational autoencoder}. The audio encoder $E$ takes mel-spectrogram $x$ as input and outputs compressed latent $z=E(x)$. The audio decoder $D$ reconstructs the mel-spectrogram signals $x'=D(z)$ from the compressed representation $z$. VAE solves the problem of excessive smoothing in mel-spectrogram reconstruction through adversarial training with a discriminator. The training objective is to minimize the weighted sum of reconstruction loss $\mathcal{L}_{re}$, GAN loss $\mathcal{L}_{GAN}$ and KL-penalty loss $\mathcal{L}_{KL}$. 
  
2) \textbf{Training latent diffusion model}. The text encoder $f_{text}$ encodes the text input $y$ into conditional embedding $c=f_{text}(y)$.  The training objective of the U-Net diffusion module is to minimize the mean squared error in the noise space:
\begin{equation}
    \label{score_loss}
    \mathcal{L}_{\theta} = \| \epsilon_\theta(z_t, t, c)-\epsilon\|_2^2,
\end{equation}
where, $\epsilon \sim \mathcal{N}(0,I)$ denotes the noise, $\epsilon_\theta$ denotes the denoising network, $t$ is the random time step, $c$ is involved by the cross-attention mechanism. The diffusion model can be efficiently trained by optimizing ELBO, ensuring extremely faithful reconstructions that match the ground-truth distribution. 

During inference, with the conditional embedding $c$ from text and the noise $z_t$ sampled from Gaussian distribution, the diffusion network runs denoising steps to get $z_0$. By applying the audio decoder $D$ and the separately trained Vocoder $V$, the generated audio is got by $a'=V(D(z_0))$. To enhance control over conditional generation, Make-An-Audio applies classifier-free guidance~\cite{ho2021classifier} technique:
\begin{equation}
\small
    \label{guidance}
    \tilde{\epsilon}_\theta(z_t, t, c) = \epsilon_\theta(z_t, t, c_{\emptyset}) + s \cdot (\epsilon_\theta(z_t, t, c) - \epsilon_\theta(z_t, t, c_{\emptyset})),
\end{equation}
where $c_{\emptyset}$ denotes the conditional embedding when empty prompt is given, $s$ is the guidance scale. The model reduces the probability of generating samples that do not use conditioning information, in favor of the samples that explicitly do as $s$ increases. 

Pseudo prompt enhancement method with audio captioning model~\cite{xu2020crnn} and audio-text retrieval model~\cite{koepke2022audio} is also introduced to construct captions for audios without natural language annotation. However, it should be noted that the quality of the captions generated by the audio captioning model and the diversity of the caption templates are limited.

\begin{figure*}[!t]
    \centering
    \includegraphics[width=\textwidth]{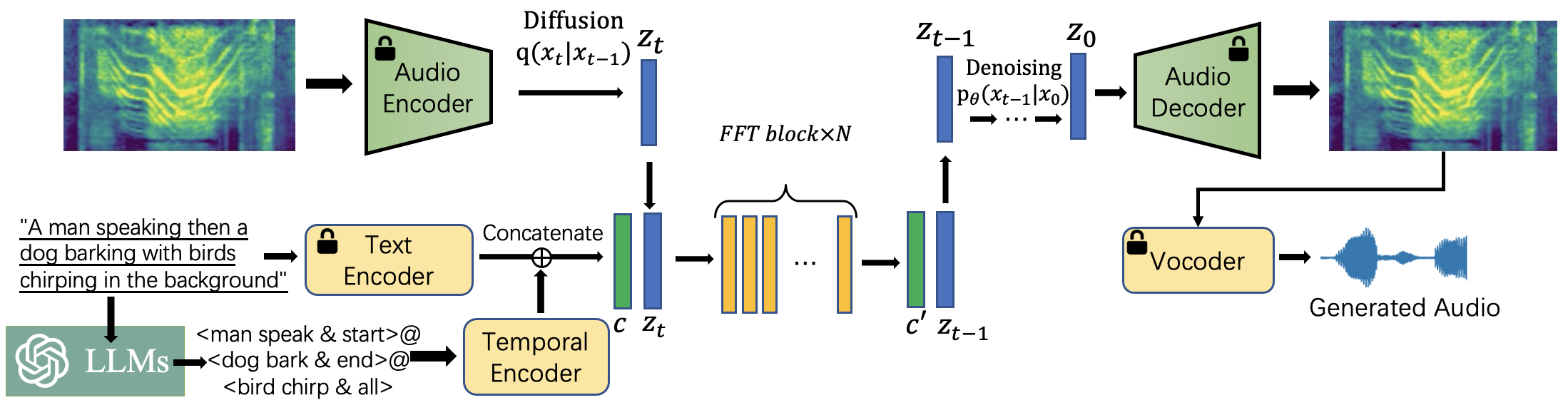}
    \caption{A high-level overview of Make-An-Audio 2. Note that modules printed with a \textit{lock} are frozen when training the T2A model.} 
    \label{fig:arch}
\end{figure*}

\begin{figure*}[!t]
    \centering
    \includegraphics[width=\textwidth]{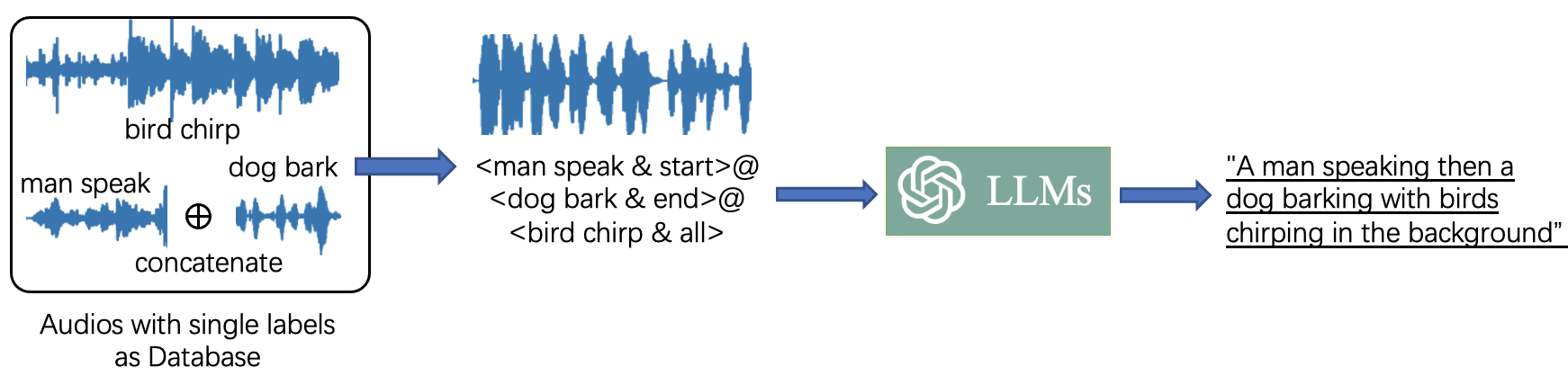}
    \caption{Overview of LLM-based data augmentation. We use single-labeled audios and their labels as a database. Composing complex audios and the structured captions with these data. We then use LLM to generate diverse natural language captions by the constructed captions and appropriate prompt.} 
    \label{fig:data_aug}
\end{figure*}
\section{Make-An-Audio 2}
In this section, we first describe the motivation of several designs in Make-An-Audio 2, and then introduce the overall architecture as illustrated in Figure~\ref{fig:arch}, to enhance temporal modeling in text-to-audio generation.

\subsection{Motivation}
In comparison to image data, audio data includes temporal information. A sound event can occur at any time within the audio, making audio synthesis a challenge when attempting to maintain temporal consistency. Previous approaches have encountered difficulties in dealing with captions that contain multiple sounds and complex temporal information, leading to semantic misalignment and poor temporal consistency. This can cause the generated audio to omit some sounds and produce an inaccurate temporal sequence. To address these issues, we propose the \textbf{temporal enhancement} method by parsing the original caption into structured pairs of <event \& order>. 

Additionally, we introduce a temporal encoder that utilizes the structured caption to help the model better understand the temporal information regarding the event sound. The temporal encoder and the main text decoder form our \textbf{dual text encoders} architecture.
Another challenge in T2A synthesis pertains to data scarcity. Most of the data in this task is "dirty", which means there are extra noises and other sounds in the audio aside from the annotated sounds. Furthermore, there is little data available with detailed temporal annotation. To combat this issue, we propose a structured data construction and \textbf{LLM-based data augmentation} approach. 

Previous diffusion-based T2A work continues the T2I diffusion model architecture using 2D-convolution and spatial transformer stacked U-Net networks as the diffusion model backbone. However, mel-spectrogram, unlike images, is not spatially translational invariant and the previous architecture is not suitable for training variable-length audio data, so we propose a feed-forward \textbf{Transformer-based diffusion denoiser backbone} and 1D-convolution-based audio VAE to improve the robustness of the model for generating variable-length audio data.

\subsection{Temporal enhancement}
Recently, AudioGPT~\cite{huang2023audiogpt} and HuggingGPT~\cite{shen2023hugginggpt} take LLM (e.g., ChatGPT) as a controller to invoke other AI models for expanding LLM's capacity in addressing multi-modal tasks. Conversely, we consider the possibility of utilizing the robust language understanding capabilities of LLMs to provide temporal knowledge. Thus, we introduce the proposed temporal enhancement method.

Specifically, LLMs are utilized to parse the input text (the natural language audio caption) and extract structured <event \& order> pairs. As illustrated in Figure~\ref{fig:arch}, LLMs simplify the original natural language caption and link each sound event to its corresponding order. Benefiting from enhanced temporal knowledge, the T2A model is empowered to identify sound events and corresponding temporal order. Appendix~D contains further details on prompt design and additional examples of temporal information enhancement.

\subsection{Dual text encoders}
To enhance the utilization of caption information, we propose a dual text encoder architecture consisting of a main text encoder CLAP~\cite{elizalde2022clap}, and a temporal encoder T5~\cite{raffel2020exploring}. With contrastive multi-modal pre-training, the CLAP has achieved excellent zero-shot performance in several downstream tasks. We freeze the weights of the main text encoder and fine-tune the temporal encoder to capture information about the temporal order of various events. 
Denote the audio-text pair and the parsed structured caption respectively as $(a,y), y_s$, the final textual representation is expressed as:
\begin{equation}
    c=Linear(Concat(f_{text}(y),f_{temp}(y_s))),
\end{equation}
where $f_{text}$ is the main text encoder and $f_{temp}$ is the temporal encoder.

\subsection{LLM-based data augmentation}
\label{LLMDA}
The remarkable success of the GPT series has underscored the significance of data-centric artificial intelligence~\cite{zha2023data,jakubik2022data}, which seeks to optimize the creation, selection, and maintenance of training and inference data for optimal outcomes. While the T2I task benefits from billions of text-image pairs~\cite{schuhmann2022laion5b}, there are currently only around one million open-source text-audio pairs avaiable~\cite{huang2023makeanaudio}. Furthermore, many of these audios are only roughly labeled with tags instead of natural language captions. To optimize the use of this data, we propose an LLM-based data augmentation technique. As shown in Figure~\ref{fig:data_aug}, we augment audio data and its corresponding text caption as follows:
\begin{itemize}[leftmargin=*]
\item We begin by collecting data labeled with single tags to create our event database $\mathcal{D}$. This type of data is typically cleaner and less likely to contain unwanted noise or other sounds. We can then use this data to construct more complex data based on their durations.
\item Then we randomly select $N \in \{2,3\}$ samples from $\mathcal{D}$, mix and concatenate them at random. Concatenating at random intervals or overlaps ensures that the resulting audio contains temporal information. Mixing improves the models' ability to recognize and separate different sorts of audio for creating complex compositions. 
\item As the resulting audio is created, we synthesize structured captions based on the occurrence time and duration of each sound event by rules. For those events that appear almost throughout the audio, we bind them with "all". While for events that only partly occur in the audio, we bind them with "start", "mid" or "end" depending on the proportion of their occurrence time points in the resulting audio. 
\item Finally, we feed the structured captions into LLM with appropriate prompts to generate diverse natural language captions. The prompt to transform structured captions to natural language captions and some examples are displayed in Appendix~D.
\end{itemize}

\subsection{Transformer-based diffusion denoiser backbone}
\label{FFT}
Previous diffusion-based work on T2A synthesis treated the mel-spectrogram as a one-channel image similar to the approach used for T2I synthesis. However, unlike images, the mel-spectrogram is not spatially translation invariant. The height of the mel-spectrogram represents the frequency domain, which means that mel-spectrogram patches at different heights can have entirely different meanings and should not be treated equally. Furthermore, the use of a 2D-convolution layer and spatial transformer-stacked U-Net architecture limits the model's ability to generate variable-length audio. Previous works~\cite{dit,uvit} has shown U-Net is not necessary for diffusion network~\cite{ho2020denoising,rombach2022high} and found transformer-based~\cite{vaswani2017attention} architecture as diffusion network can achieve better performance. Inspired by these works and to improve the model's ability to generate variable-length audio, we propose a modified audio VAE that uses a 1D-convolution-based model and propose a feed-forward Transformer-based diffusion denoiser backbone. While the latent of Make-An-Audio's audio encoder is $z=E(x)\in R^{c_e\times C_a/f\times T/f}$, where $c_e$ is the embedding dim of latent, $f$ is the downsampling rate, $C_a$ and $T$ denote the mel-channels and the number of frames of mel-spectrogram respectively, which can be seen as images' height and width. Our 1D-convolution-based audio encoder's latent is  $z=E(x)\in R^{C_a/f_1\times T/f_2}$, where $f_1,f_2$ are downsampling rates of mel-channels and frames, respectively. 
As the feed-forward Transformer block is composed of 1D-convolution and temporal transformer, it can better understand the information of the temporal domain in latent and improves the variable length audio generation performance. Compared to the original spatial transformer, the computation complexity reduces from $O((C_a/f \times T/f)^2 \times D)$ to $O((T/f_2)^2 \times D)$, where $D$ is the embedding dimension of the transformer layer.

\section{Experiments}

\subsection{Experimental setup}
\paragraph{Dataset.} \label{data}
We use a combination of several datasets to train our model, including: AudioCaps training set, WavCaps, AudioSet, Adobe Audition Sound Effects, Audiostock,  ESC-50, FSD50K, MACS, Epidemic Sound, UrbanSound8K, WavText5Ks, TUT acoustic scene. 
This results in a dataset composed of \textbf{0.92 million} audio text pairs, with a total duration of approximately \textbf{3.7K hours}.

We conduct preprocessing on both text and audio data as follows:
1) We convert the sampling rate of audios to 16kHz. Prior works~\cite{yangDiffsoundDiscreteDiffusion2023,huang2023makeanaudio,liuAudioLDMTexttoAudioGeneration2023} pad or truncate the audio to a fixed length (10s), while we group audio files with similar durations together to form batches to avoid excessive padding which could potentially impair model performance and slow down the training speed. This approach also allows for improved variable-length generation performance. We truncate any audio file that exceeds 20 seconds, in order to speed up the training process. 2) We adopt the LLM-based data augmentation method in section~\ref{LLMDA} to construct approximately 61k additional audio-text pairs as auxiliary data. 3) For audios without natural language annotation, we apply the pseudo prompt enhancement method from Make-An-Audio~\cite{huang2023makeanaudio} to construct captions aligned with the audio. 4) We assign a lower weightage to the data that is not annotated with temporal information, but is abundant in quantity and diversity, such as AudioSet and WavCaps data. Specifically, we traverse the AudioCaps training set and the LLM augmented data with a probability of 50\%, while randomly selecting data from all other sources with a probability of 50\%. For the latter dataset, we use "<text \& all>" as their structured caption. 
To evaluate the performance of our models, we use the AudioCaps test set and Clotho evaluation set which contain multiple event audio samples and detailed audio captions. The latter serves as a more challenging zero-shot scenario test for us. More details about datasets are put in Appendix~A.

\paragraph{Model configurations.}
We train a continuous 1D-convolution-based VAE to compress the mel-spectrogram to a 20-channel latent with temporal axis downsample rates of 2. For our main experiments, we train a feed-forward Transformer-based diffusion backbone. The diffusion model is trained on 8 NVIDIA A100 GPU with 1.2M optimization steps, 4 batch size per GPU. We use the AdamW optimizer~\cite{kingma2014adam} with a learning rate of 9.6e-5. BigVGAN~\cite{leeBigVGANUniversalNeural2023}, a universal vocoder that generalizes well for various scenarios, is used as our vocoder, we trained it on the AudioSet Dataset. For more details about Model configuration please refer Appendix~B.

\paragraph{Evaluation methods.}  \label{training}
We evaluate our models using objective and subjective metrics to assess the audio quality and text-audio alignment faithfulness. For objective evaluation, we include Frechet distance (FD), inception score (IS), Kullback–Leibler (KL) divergence, Frechet audio distance (FAD), and CLAP score. For subjective evaluation, we conduct crowd-sourced human evaluations with MOS (mean opinion score) to assess the audio quality and the text-audio alignment faithfulness, scoring MOS-Q and MOS-F, respectively. More information regarding the evaluation process can be found in Appendix~C.2 

\paragraph{Baseline models.}
To establish a standard for comparison, our study employs four baseline models, including Make-An-Audio~\cite{huang2023makeanaudio}, AudioLDM~\cite{liuAudioLDMTexttoAudioGeneration2023}, TANGO~\cite{ghosal2023texttoaudio} and AudioGen~\cite{kreukAudioGenTextuallyGuided2023}. We reimplement Make-An-Audio and train it on AudioCaps Dataset. AudioLDM-S and AudioLDM-L with 454M and 1.01B parameters respectively are trained on AudioCaps, AudioSet, BBC Sound Effects, and the FreeSound dataset. TANGO is trained on AudioCaps Dataset.  The close-source model AudioGen-Base and AudioGen-Large, with 285M and 1M parameters are trained on AudioCaps, AudioSet, and eight other datasets.

\subsection{Main results}
\textbf{Automatic objective evaluation.} 
The objective evaluation comparison with baseline is presented in Table~\ref{tab:comp}, and we have the following observations: 1) In terms of audio quality, Make-An-Audio 2 achieves better scores in FD, IS, KL, and FAD; 2) On text-audio similarity, Make-An-Audio 2 presents the comparable score in CLAP; 3) Regarding temporal alignment, Make-An-Audio 2 also achieves a high IS score, showing that temporal enhancement method also enhances the clarity and expressiveness of the sound events in the generated audio.



\textbf{Subjective human evaluation.}
The human evaluation results show significant gains of Make-An-Audio 2 with MOS-Q of 78.31 and MOS-F of 75.63, outperforming TANGO in the MOS-F evaluation. It indicates that raters prefer our model synthesis against baselines in terms of audio naturalness and faithfulness.

\textbf{Zero-shot evaluation.} 
To further investigate the generalization performance of the models, we additionally test the performance of the models on the Clotho-evaluation dataset in the zero-shot scenario. As illustrated in Table~\ref{tab:clotho_5scaps_result}, Make-An-Audio 2 has significantly better results than TANGO and AudioLDM-L, attributing to the scalability in terms of data usage.

\begin{table*}[]
\centering
\caption{The comparison between Make-An-Audio 2 and baseline T2A models on the AudioCaps dataset. All the diffusion-based models run with 100 DDIM~\cite{ddim} steps for a fair comparison. We borrowed all the results from~\cite{liuAudioLDMTexttoAudioGeneration2023,ghosal2023texttoaudio} and used the model released by the authors on Huggingface to test CLAP Score. We reimplement Make-An-Audio and replace their vocoder with our BigVGAN vocoder. }
\vspace{2mm}
\small
\begin{tabular}{ccccccccc}
\toprule
Model  & Prams  & FD↓    & IS↑    & KL↓    & FAD↓   & CLAP↑  & MOS-Q↑ & MOS-F↑ \\
\midrule
GroundTruth & -   & -      & -      & -   & -   & 0.671   &  86.47     & 84.31  \\
AudioGen-S & 285M   & -      & -      & 2.09   & 3.13   & -      & -      & -  \\
AudioGen-L & 1B     & -      & -      & 1.69   & 1.82   & -      & -      & -  \\
Make-An-Audio & 453M   & 18.32  & 7.29   & 1.61   & 2.66   & 0.593  &  69.54    & 65.45  \\
AudioLDM-S & 454M   & 29.48  & 6.9    & 1.97   & 2.43   &  -     & -      & - \\
AudioLDM-L & 1.01B   & 23.31  & 8.13   & 1.59   & 1.96   & 0.605   &  70.91      & 67.41 \\
TANGO  & 1.21B   & 26.13  & 8.23   & 1.37   & 1.87   & $\mathbf{0.650}$  &  74.10   & 72.76 \\
\midrule
Make-An-Audio 2 & 937M   & $\mathbf{11.75}$  & $\mathbf{11.16}$  & $\mathbf{1.32}$    & $\mathbf{1.80}$   &  0.645      &   $\mathbf{78.31}$     & $\mathbf{75.63}$ \\
\bottomrule
\end{tabular}%
\label{tab:comp}%
\end{table*}%

\begin{table*}[!h]
  \centering
  \small
  \caption{Comparison of Make-An-Audio 2, AudioLDM-L, and Tango on Clotho-eval (10s) and AudioCaps-test (5s) datasets. We truncate the ground truth audios from AudioCaps dataset to 5 seconds to conduct variable-length generation evaluation.}
  \vspace{2mm}
  \begin{tabular}{c|cccc|ccccc}
    \toprule
    \multirow{2}{*}{Model} & \multicolumn{4}{c|}{Clotho-eval} & \multicolumn{4}{c}{AudioCaps-test} \\
    & FD↓ & IS↑ & KL↓ & FAD↓ & FD↓ & IS↑ & KL↓ & FAD↓ \\
    \midrule
    TANGO & 32.1 & 6.77 & 2.59 & 3.61 & 31.76 & 5.50 & 2.04 & 10.53 \\
    AudioLDM-L & 28.15 & 6.55 & 2.6 & 4.93 & 31.97 & 5.66 & 2.39 & 6.79 \\
    Make-An-Audio 2 & \textbf{19.97} & \textbf{8.50} & \textbf{2.49} & \textbf{2.13} & \textbf{13.78} & \textbf{9.95} & \textbf{1.61} & \textbf{2.33} \\
    \bottomrule
  \end{tabular}
  \label{tab:clotho_5scaps_result}
\end{table*}

\subsection{Analyses}

\textbf{Variable-length generation.} Audio data can have different lengths, to investigate our models' performance on variable-length audio generation, we test to generate 5 seconds audios on AudioCaps dataset, the results are shown in \autoref{tab:clotho_5scaps_result}. From the table, it can be seen that TANGO and AudioLDM exhibit significant performance degradation when generating audio with different lengths than the training data, as TANGO and AudioLDM pad or truncate all the training audio data to 10 seconds, and their models are based on 2D-convolution and spatial transformer to process mel-spectrogram as images. Make-An-Audio 2 maintains high performance even when generating variable-length audio samples since it is trained on audio samples of varying lengths and utilizes 1D-convolution and temporal transformers to emphasize temporal information.

\begin{table}[!h]
\centering
\small
\caption{Comparison of different classifier-free guidance scales on AudioCaps-test set.}
\label{tab:guidance}%
\begin{tabular}{ccccc}
\toprule
Guidance Scale & FD↓    & IS↑    & KL↓    & FAD↓ \\
\midrule
1      & 30.13  & 4.61   & 2.26   & 7.55 \\
2      & 20.71  & 7.23   & 1.66   & 3.76 \\
3      & 15.71  & 9.07   & 1.47   & 2.41 \\
4      & \textbf{11.75}  & 11.16  & 1.32   & \textbf{1.80} \\
5      & 12.43  & \textbf{11.37}  & \textbf{1.25}   & 1.95 \\
7.5    & 13.96  & 11.32  & 1.31   & 2.97 \\
10     & 16.06  & 10.52  & 1.47   & 3.59 \\
\bottomrule
\end{tabular}%
\end{table}%

\textbf{Classifier-free guidance scale.}
We explore the classifier-free guidance in text-to-audio synthesis. The choice of the classifier guidance scale could scale conditional and unconditional synthesis, offering a trade-off between sample faithfulness and realism with respect to the conditioning text. We show the effect of guidance scale $w$ on T2A in \autoref{tab:guidance}. Our results show that a guidance scale of $w=4$ yields the best performance in terms of FD and FAD.

\begin{table}[!h]
\centering
\small
\caption{Vocoder reconstruction performance comparison on AudioCpas-test set. }
\begin{tabular}{cccccc}
\toprule
  Vocoder  & Model    & FD↓    & IS↑    & KL↓    & FAD↓ \\
\midrule
MelGAN  &  Diffsound & 26.14  & 5.4    & 1.22   & 6.24 \\
HifiGAN  &  Make-An-Audio & 21.79  & 5.93   & 1.03   & 6.02 \\
HifiGAN  &  AudioLDM & 11.45  & 8.13   & 0.22   & 1.18 \\
BigVGAN & Make-An-Audio 2 & 5.45   & 9.44   & 0.17   & 0.98 \\
\bottomrule
\end{tabular}%
\label{tab:vocoder}
\end{table}%

\textbf{Vocoder performance comparison.}
\label{vocoder_comp}
Vocoder is another key component to improve the generated audio quality for mel-spectrogram-based models. We further compared the 1) BigVGAN vocoder used in Make-An-Audio 2, 2) AudioLDM's pre-trained HifiGAN, 3) Make-An-Audio's pre-trained HifiGAN and 4) Diffsound's pre-trained MelGAN. The results are shown in \autoref{tab:vocoder}. We find 3) and 4) both perform worse compared with 2), while 2) and 3) use the same vocoder architecture. We find that the problem lies in the mel-processing method of 3) and 4), they use the same mel-processing method which will result in poor performance of the vocoder. So we adopt the same mel-spectrogram extraction process as BigVGAN and get the best performance.

\subsection{Ablation study}

\begin{table}[!h]
\centering
\caption{The results evaluated on AudioCaps-test set with different settings of text encoder, diffusion model backbone, and dataset. Noted that $^{\dagger}$ marks the text encoder is frozen, $^{*}$ means trainable one.}
\label{tab:ablation}%
  \small
    \begin{tabular}{c|lcccccc}
    \toprule
    Setting & Text encoder & Backbone & Dataset & FD↓    & IS↑    & KL↓    & FAD↓ \\
    \midrule
    \#1 & CLAP$^{\dagger}$ & U-Net  & AudioCaps & 18.24  & 7.46   & 1.61   & 2.25 \\
    \#2 & T5$^{*}$  & U-Net  & AudioCaps & 13.73  & 9.79   & 1.41   & 2.05 \\
    \#3 & CLAP$^{\dagger}$+T5$^{*}$ & U-Net  & AudioCaps & 13.71  & 9.99   & 1.36   & 1.92 \\
    \#4 & CLAP$^{\dagger}$+T5$^{*}$ & U-Net  & Variable length All & 22.69  & 5.93   & 2.17   & 3.82 \\
    \#5 & CLAP$^{\dagger}$+T5$^{*}$ & Transformers & Variable length All & 11.75  & 11.16  & 1.32   & 1.80 \\
    \bottomrule
    \end{tabular}%
\end{table}%

In order to assess the effectiveness of various designs in Make-An-Audio 2 including temporal enhancement, dual text encoder, and feed-forward Transformer-based diffusion denoiser, we conduct ablation studies. The results evaluated on the AudioCaps-test set with different settings for text encoder, diffusion denoiser backbone, and dataset are presented in \autoref{tab:ablation}. The key findings are discussed below:
\begin{itemize}[leftmargin=*]
    \item Comparing setting \#2 to \#1 highlights the effectiveness of temporal enhancement. We use LLM to extract event and temporal information and create structured input in the form of <event \& order> pairs. As this format of structured input is not in the training corpus of the text encoder, we use the trainable T5 as our text encoder, which leads to significant improvements in both objective scores and sound timing modeling.
    \item Comparing setting \#3 with \#2 proves the effectiveness of dual text encoder architecture. The reason is that the structured caption parsed by LLM may result in the loss of some information such as adjectives and quantifiers, and the trainable text encoder may reduce generalization abilities. Therefore, it is still essential to use the frozen CLAP encoder to comprehend the original natural language caption.
    \item For setting \#4, we train our model with the large combined dataset instead of AudioCaps-train set and switch the training mode from fixed-length audio training to variable-length audio, resulting in severe performance degradation. As 2D-convolution and spatial Transformer treat the mel-spectrograms as images, and it treats the temporal information and frequency information the same way, they demonstrate deficiencies in understanding variable-length data with temporal information. 
    \item Setting \#5 is our Make-An-Audio 2. Compared with setting \#4, we replace the 2D-convolutional VAE with 1D-convolutional VAE and replace the U-Net diffusion backbone with feed-forward Transformer layers and observe performance gains, which demonstrates a strong ability in understanding variable length audios and temporal information.
\end{itemize}

\section{Conclusions}
In this work, we present Make-An-Audio 2, a temporal-enhanced T2A synthesis model. With a capable LLM to extract temporal information from the natural language caption, Make-An-Audio 2 can better understand the event order in the caption and generate semantically aligned audios. Leveraging 1D-convolutional VAE and feed-forward Transformer diffusion backbone, Make-An-Audio 2 can generate variable-length audios without performance degeneration. With complex audio reconstruction and LLM-based data augmentation, Make-An-Audio 2 is endowed with the ability to understand complex temporal relationships and combinations of multiple concepts. Make-An-Audio 2 achieves the SOTA audio generation quality in both objective and subjective metrics, and extends the boundaries of the T2A. We discuss the limitations, future works and broader impact in Appendix~E.
\bibliographystyle{plain}
\bibliography{neurips_2023}
\newpage

\newpage
\appendix
\onecolumn

\begin{center}{\bf {\LARGE Appendices} }
\end{center}
\begin{center}{\bf {\Large Make-An-Audio 2: Temporal-Enhanced Text-to-Audio Generation} \linebreak}
\end{center}
\section{Data details}  \label{app:data}

\begin{table}[htbp]
  \centering
    \begin{tabular}{llll}
    \toprule
    Dataset & Hours  & Type   & Source \\
    \midrule 
    Audiocaps & 109hrs & caption & ~\cite{kim2019audiocaps} \\
    WavCaps & 2056hrs & caption & ~\cite{mei2023wavcaps} \\
    WavText5K & 25hrs  & caption & ~\cite{deshmukh2022audio} \\
    MACS   & 48hrs  & caption & ~\cite{martin2021ground} \\
    Clothv2 & 152hrs & caption & ~\cite{drossos2020clotho} \\
    Audiostock & 44hrs  & caption &  \url{https://audiostock.net} \\
    epidemic sound & 220hrs & caption &  \url{https://www.epidemicsound.com} \\
    Adobe Audition Sound Effects & 26hrs  & caption  & \multicolumn{1}{p{19.085em}}{\url{https://www.adobe.com/products/audition/offers/AdobeAuditionDLCSFX.html}} \\
    \midrule 
    FSD50K & 108hrs & label  & \url{https://annotator.freesound.org/fsd} \\
    ODEON\_Sound\_Effects & 20hrs  & label  & \multicolumn{1}{p{19.085em}}{\url{https://www.paramountmotion.com/odeon-sound-effects}} \\
    UrbanSound8K & 9hrs   & label  & ~\cite{Salamon:UrbanSound:ACMMM:14} \\
    ESC-50 & 3hrs   & label  & ~\cite{piczak2015esc} \\
    filteraudioset & 945hrs & multi label & ~\cite{gemmeke2017audio} \\
    TUT    & 13hrs  & label  & ~\cite{TUT} \\
    \bottomrule
    \end{tabular}%
  \vspace{5pt}
  \caption{Statistics for the Datasets used in the paper.}
  \label{tab:dataset}%
\end{table}%

As shown in Table~\ref{tab:dataset}, we collect a large-scale audio-text dataset consisting of 0.92 million of audio samples with a total duration of approximately 3.7k hours. The dataset has a wide variety of sounds including music and musical instruments, sound effects, human voices, nature and living sounds, etc. For Clotho dataset, we only use its evaluation set for zero-shot testing and do not use for training. As speech and music are the dominant classes in AudioSet, we filter 95\% of the samples that contain speech and music to build a more balanced dataset.

\section{Experimental details}  \label{app:expeiment}
\paragraph{Variational autoencoder.}
 We employed a similar VAE architecture to that of Make-An-Audio, replacing all the 2D-convolution layers with 1D-convolution layers and the spatial transformer with a temporal transformer. As detailed in Section~4.5,  
 the output latent of VAE is $z=E(x)\in R^{C_a/f_1\times T/f_2}$, where we choose the downsample rate of $f_1=4$ and $f_2=2$. We additionally involve R1 regularization~\cite{r1reg} to better stabilize the adversarial training process. 
We train our VAE on 8 NVIDIA A100 GPU with a batch size of 32 and 800k training steps on AudioSet dataset. We use the Adam optimizer~\cite{kingma2014adam} with a learning rate of $1.44\times 10^{-4}$. For specific differences in hyperparameters between our VAE and that of Make-An-Audio, please see Table~\ref{tab:VAE_diff}.

\begin{table}[htbp]
  \centering
    \begin{tabular}{c|c|c}
    \toprule
      & Make-An-Audio VAE & Make-An-Audio 2 VAE \\
    \midrule 
    \multicolumn{1}{c|}{Assume input tensor shape (for 10s audio)} & (1,80,624) & (80,624) \\
    Embed\_dim & 4      & 20 \\
    Convolution layer & Conv2D & Conv1D \\ 
    Channels & 128    & 224 \\
    Channel multiplier & 1,2,2,4 & 1,2,4 \\
    Downsample layer position & after block 1,2 & after block 1 \\
    Attention layer & spatial attention & temporal attention \\
    Attention layer position & after block 3,4 & after block 3 \\
    Output tensor shape & (4,10,78) & (20,312) \\
    \bottomrule
    \end{tabular}%
  \vspace{5pt}
  \caption{Difference between Make-An-Audio VAE and our VAE}
  \label{tab:VAE_diff}%
\end{table}%

\paragraph{Latent diffusion.}
We train our Latent Diffusion Model with on 8 NVIDIA A100 GPU with a batch size of 32 and 1.2M training steps. We use the Adam optimizer with a learning rate of $9.6\times 10^{-5}$. For the specific hyperparameter for our latent diffusion model, please refer to Table~\ref{tab:LDM_config}.

\begin{table}[htbp]
  \centering
    \begin{tabular}{c|c}
    \toprule
           & Make-An-Audio 2 LDM \\
    \midrule 
    Input shape & (20,T) \\
    Condition\_embedding dim & 1024 \\
    Feed-forward Transformer hidden\_size & 576 \\
    Feed-forward Transformer's Conv1d kernel size & 7 \\
    Feed-forward Transformer's Conv1d padding & 3 \\
    Number of Transformer heads & 8 \\
    Number of Feed-forward Transformer block & 8 \\
    Diffusion steps & 1000 \\
    \bottomrule
    \end{tabular}%
  \vspace{5pt}
  \caption{Make-An-Audio 2 Diffusion model backbone configurations}
  \label{tab:LDM_config}%
\end{table}%

\paragraph{Model parameters of each component.}
The params of each component in Make-An-Audio 2 are displayed in Table~\ref{tab:component_param}.

\begin{table}[htbp]
  \centering
    \begin{tabular}{cc}
    \toprule
    Component & Params \\
    \midrule 
    VAE    & 213M \\
    Diffusion Model Backbone & 160M \\
    Text Encoder & 452M \\
    Vocoder & 112M \\
    \midrule 
    Total  & 937M \\
    \bottomrule
    \end{tabular}%
  \vspace{5pt}
  \caption{The params of each component in Make-An-Audio2}
  \label{tab:component_param}%
\end{table}%

\section{Evaluation} 
\subsection{Subjective evaluation}  \label{app:human_eval}
To assess the generation quality, we conduct MOS (Mean Opinion Score) tests regarding audio quality and text-audio faithfulness, respectively scoring MOS-Q and MOS-F. 

For audio quality, the raters were explicitly instructed to ``focus on examining the audio quality and naturalness.'' The testers were presented with audio samples and asked to rate their subjective score (MOS-P) on a 20-100 Likert scale. 

For text-audio faithfulness, human raters were shown the audio and its caption and asked to respond to the question, "Does the natural language description align with the audio faithfully?" They had to choose one of the options - "completely," "mostly," or "somewhat" on a 20-100 Likert scale. 

Our crowd-sourced subjective evaluation tests were conducted via Amazon Mechanical Turk where participants were paid \$8 hourly. A small subset of the generated audio samples used in the test can be found at \url{https://make-an-audio-2.github.io/}.

\subsection{Objective evaluation}  \label{app:obj_eval}
Fréchet Audio Distance (FAD)~\cite{kilgour2019frechet} is adapted from the Fréchet Inception Distance (FID) to the audio domain, it is a reference-free perceptual metric that measures the distance between the generated and ground truth audio distributions. FAD is used to evaluate the quality of generated audio.

The inception Score (IS) is an effective metric that evaluates both the quality and diversity of generated audio. 

KL divergence is measured at a paired sample level between the generated audio and the ground truth audio, it is computed using the label distribution and is averaged as the final result. 

Fréchet Distance (FD) evaluates the similarity between the generated and ground truth audio distributions. FD, KL and IS are built upon an audio classifier, PANNs~\cite{kong2020panns}, which takes the mel-spectrogram as model input. Differently, FAD uses VGGish~\cite{vggish_hershey2017cnn} as an audio classifier that takes raw audio waveform as model input. 

CLAP score: adapted from the CLIP score~\cite{hessel2021clipscore,radford2021learning} to the audio domain and is a reference-free evaluation metric to measure audio-text alignment for this work that closely correlates with human perception.

\section{ChatGPT prompts}  \label{app:gpt_prompt}
\begin{figure*}[!t]
    \centering 
    \includegraphics[width=\textwidth]{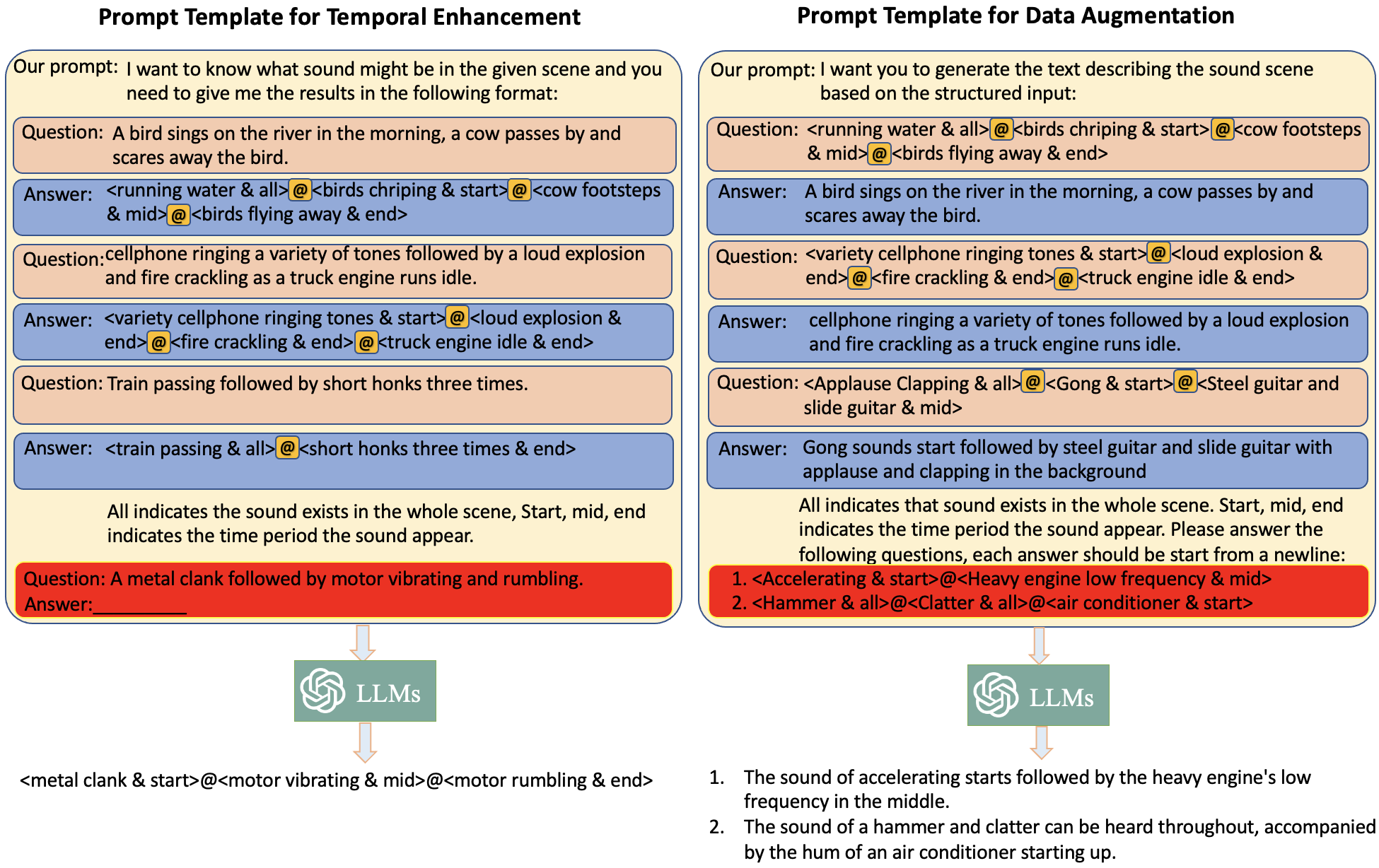}
    \caption{The prompt templates we used for temporal enhancement and data augmentation. We use the symbol '\&' to split the sound event and the time order. We use the symbol '@' to split <event \& order> pairs.} 
    \label{fig:prompts}
\end{figure*}

The prompt templates utilized for temporal enhancement to construct structure caption from the original natural language caption  and for caption data augmentation are displayed in Figure~\ref{fig:prompts}.

Table~\ref{tab:extract_example} presents some instances of the original caption and ChatGPT's outcome. For text data augmentation, we construct structured caption inputs, and Table~\ref{tab:data_aug} exhibits examples of such inputs and ChatGPT's corresponding outputs.

\begin{table}[htbp]
  \centering
    \begin{tabular}{p{14.835em}|p{22.335em}}
    \hline
    Natural language input & ChatGPT's output \\
    \hline
    A woman talks nearby as water pours & <woman talking\& all>@<water pouring\& all> \\
    \hline
    Two men exchange words, then a car engine revs followed by a siren and fade in music & <two men talking\& all>@<car engine revving\& start>@<siren\& mid>@<music fading in\& end> \\
    \hline
    A crowd is cheering and shouting, thumping occurs, an adult female speaks, and an adult male speaks & <crowd cheering and shouting\& all>@<thumping\& start>@<adult female speaking\& mid>@<adult male speaking\& end> \\
    \hline
    \end{tabular}%
  \vspace{5pt}
  \caption{Examples of using ChatGPT for temporal enhancement from AudioCaps trainset}
  \label{tab:extract_example}%
\end{table}%

\begin{table}[htbp]
  \centering
    \begin{tabular}{p{14.835em}|p{22.335em}}
    \hline
    Structured input & ChatGPT's output \\
    \hline
    <bark howl dog\& start>@<typing Typewriter\& mid>@<breathing\& end> & A dog barks and howls while someone types on a typewriter, then the sound of breathing takes over \\
    \hline
    <crowing\& all>@<car\& all>@<female singing\& end> & A rooster crows as a car passes by, and the scene ends with the sound of a female singing \\
    \hline
    <sneezing\& all>@<bicycle bell ring\& start>@<typewriter \& end> & The sound of sneezing is heard throughout, with a bicycle bell ringing at the start and the sound of a typewriter at the end \\
    \hline
    \end{tabular}%
  \vspace{5pt}
  \caption{Examples of using ChatGPT for data augmentation}
  \label{tab:data_aug}%
\end{table}%

\section{Limitations, future works and broader impact}  \label{app:limit}
\paragraph{Limitations.}
Make-An-Audio 2 incorporates an additional LLM for parsing the original caption, which affects both the generation performance and running speed. Additionally, the generative diffusion model employed by Make-An-Audio 2 requires multiple iterative refinements for synthesis, which can be time-consuming to produce high-quality results. Furthermore, the speech in the generated audio can be intelligible.
\paragraph{Future works.}
We leave the T2A system which supports speech synthesis for future work. In addition, we aim to implement T2A systems that could take structured inputs as optional auxiliary inputs instead of required inputs.
\paragraph{Broader impacts.}
At the same time, we acknowledge that Make-An-Audio 2 may lead to unintended consequences such as increased unemployment for individuals in related fields such as sound engineering and radio hosting. Furthermore, there are potential concerns regarding the ethics of non-consensual voice cloning or the creation of fake media.

\end{document}